\documentclass
[tightenlines,final,letterpaper,10pt,pre,amsfonts,amssymb,twocolumn,showkeys,byrevtex,preprint,noshowpacs]{revtex4}%
\usepackage{amsfonts}
\usepackage{amsmath}
\usepackage{amssymb}
\usepackage{graphicx}%
\providecommand{\U}[1]{\protect\rule{.1in}{.1in}}

\begin{document}
\title{Model of human collective decision-making in complex environments}
\author{Giuseppe Carbone$^{1,2}$, Ilaria Giannoccaro$^{1}$ }
\affiliation{$^{1}$Department of Mechanics Mathematics and Management, Politecnico di Bari,
v.le Japigia 182, 70126 Bari, Italy}
\affiliation{$^{2}$Physics Department M. Merlin, CNR Institute for Photonics and
Nanotechnologies U.O.S. Bari, via Amendola 173, 70126 Bari, Italy}

\begin{abstract}
A continuous-time Markov process is proposed to analyze how a group of humans
solves a complex task, consisting in the search of the optimal set of
decisions on a fitness landscape. Individuals change their opinions driven by
two different forces: (i) the self-interest, which pushes them to increase
their own fitness values, and (ii) the social interactions, which push
individuals to reduce the diversity of their opinions in order to reach
consensus. Results show that the performance of the group is strongly affected
by the strength of social interactions and by the level of knowledge of the
individuals. Increasing the strength of social interactions improves the
performance of the team. However, too strong social interactions slow down the
search of the optimal solution and worsen the performance of the group. In
particular, we find that the threshold value of the social interaction
strength, which leads to the emergence of a superior intelligence of the
group, is just the critical threshold at which the consensus among the members
sets in. We also prove that a moderate level of knowledge is already enough to
guarantee high performance of the group in making decisions.

\end{abstract}
\maketitle

\section{Introduction\label{introduction}}

The ability of groups to solve complex problems that exceed individual skills
is widely recognized in natural, human, and artificial contexts. Animals in
groups, e.g. flocks of birds, ant colonies, and schools of fish, exhibit
collective intelligence when performing different tasks as which direction to
travel in, foraging, and defence from predators \cite{Nature1, Nature2}.
Artificial systems such as groups of robots behaving in a self organized
manner show superior performance in solving their tasks, when they adopt
algorithms inspired by the animal behaviors in groups \cite{Nature3, Science1,
Science2, Review-Swarm-Robotics}. Human groups such as organizational teams
outperform the single individuals in a variety of tasks, including problem
solving, innovative projects, and production issues \cite{CollDM, JIGSAW,
KnowledgeTransf7,Galam1, Galam2}.

The superior ability of groups in solving tasks originates from collective
decision making: agents (animals, robots, humans) make choices, pursuing their
individual goals (forage, survive, etc.) on the basis of their own knowledge
and amount of information (position, sight, etc.), and adapting their behavior
to the actions of the other agents. The group-living enables social
interactions to take place as a mechanism for knowledge and information
sharing \cite{KnowledgeTransf7, KnowledgeTransf1, KnowledgeTransf2,
KnowledgeTransf3, KnowledgeTransf4, KnowledgeTransf5, KnowledgeTransf6,
KnowledgeTransf8, West, knowF1, knowF2}. Even though the single agents posses
a limited knowledge, and the actions they perform usually are very simple, the
collective behavior, enabled by the social interactions, leads to the
emergence of a superior intelligence of the group. This property is known as
swarm intelligence \cite{Nature4, SI, Grigolini} and wisdom of crowds
\cite{wisdom}.

In this paper we focus on human groups solving complex combinatorial problems.
Many managerial problems including new product development, organizational
design, and business strategy planning may be conceived as problems where the
effective combinations of multiple and interdependent decision variables
should be identified \cite{levinthal, katila,loch,billinger}. We develop a
model of collective decision making, which attempts to capture the main
drivers of the individual behaviors in groups, i.e., self-interest and
consensus seeking. We consider that individuals make choices based on rational
calculation and self-interested motivations. Agent's choices are made by
optimizing the perceived fitness value, which is an estimation of the real
fitness value based on the level of agent's knowledge \cite{PRE1, Larissa,
Nature1}. However, any decision made by an individual is influenced by the
relationships he/she has with the other group members. This social influence
pushes the individual to modify the choice he/she made, for the natural
tendency of humans to seek consensus and avoid conflict with people they
interact with \cite{consenso}.

We use the Ising-Glauber dynamics \cite{castellano, Glauber}\ to model the
social interactions among group members. The $N-K$ model \cite{NK-model,
NK-model1} is employed to build the fitness landscape associated with the
problem to solve. A continuous-time Markov chain governs the decision-making
process, whose complexity is controlled by the parameter $K$. We define the
transition rate of individual's opinion change as the product of the
Ising-Glauber rate (\cite{Glauber}), which implements the consensus seeking
\cite{IsingOriginal, IsingOr2, IS1, wedlich1}, and an exponential rate
\cite{weidlich2, Sweitzer} ,which speeds up or slows down the change of
opinion, to model the rational behavior of the individual.

Herein, we explore how both the strength of social interactions and the level
of knowledge of the members influence the group performance. We identify in
which circumstances human groups are particularly effective in solving complex
problems. We extend previous studies highlighting the efficacy of collecting
decision making in presence of a noisy environment \cite{Math-Imp}, and in
conditions of cognitive limitations\cite{Nature2, KnowledgeTransf7, psi1,
psi2, psi3}. This decision-making model might be proposed as optimization
technique belonging to the class of swarm intelligence techniques
\cite{Nature4, DorigoOr1, DorigoOr2, PSO, ABC, FSA}.

\section{The Model}

We consider a human group made of $M$ socially interacting members, which is
assigned to solve a complex task. The task consists in solving a combinatorial
decision making problem by identifying the set of decisions (choice
configuration) with the highest fitness. The fitness function is built
employing the $N-K$ model \cite{NK-model, NK-model1, NK-model2}. A
$N$-dimensional vector space of decisions is considered, where each choice
configuration is represented by a vector $\mathbf{d=}$ $\left(  d_{1}%
,d_{2},...,d_{N}\right)  $. Each decision is a binary variable that may take
only two values $+1$ or $-1$, i.e. $d_{i}=\pm1$, $i=1,2,...,N$. The total
number of decision vectors is therefore $2^{N}$. Each vector $\mathbf{d}$ is
associated with a certain fitness value $V\left(  \mathbf{d}\right)  $
computed as the weighted sum of $N$ stochastic contributions $W_{j}\left(
d_{j},d_{1}^{j},d_{2}^{j},..,d_{K}^{j}\right)  $, each decision leads to total
fitness depending on the value of the decision $d_{j}$ itself and the values
of other $K$ decisions $d_{i}^{j}$, $i=1,2,...,K$. Following the classical
$N-K$ \cite{NK-model, NK-model1, NK-model2} procedure (more details are
provided in Appendix \ref{NK-model}), the quantities $W_{j}\in\left[
0,1\right]  $ are determined as randomly generated $2^{K+1}$-element
\textquotedblleft interaction tables\textquotedblright. The fitness function
of the group is defined as
\begin{equation}
V\left(  \mathbf{d}\right)  =\frac{1}{N}\sum_{j=1}^{N}W_{j}\left(  d_{j}%
,d_{1}^{j},d_{2}^{j},..,d_{K}^{j}\right)  \label{fitness function}%
\end{equation}
The integer index $K=0,1,2,...,N-1$ is the number of interacting decision
variables, and tunes the complexity of the problem. The complexity of the
problem increases with $K$. Note that, for $K>2$, in computational complexity
theory, finding the optimum of the fitness function $V\left(  \mathbf{d}%
\right)  $ is classified as a NP-complete decision problem \cite{NK-model2}.
This makes this approach particularly suited in our case.

We model the level of knowledge of the $k$-th member of the group (with
$k=1,2,...M$) by defining the $M\times N$ competence matrix $\mathbf{D}$,
whose elements $D_{kj}$ take the value $D_{kj}=1$ if the member $k$ knows the
contribution of the decision $j$\textit{ }to the total fitness $V$, otherwise
$D_{kj}=0$. Based on the level of knowledge each member $k$ computes his/her
own perceived fitness (self-interest) as%
\begin{equation}
V_{k}\left(  \mathbf{d}\right)  =\frac{\sum_{j=1}^{N}D_{kj}W_{j}\left(
d_{j},d_{1}^{j},d_{2}^{j},..,d_{K}^{j}\right)  }{\sum_{j=1}^{N}D_{kj}}.
\label{perceived fitness}%
\end{equation}
Each member of the group makes his/her choices driven by the rational
behavior, which pushes him/her to increase the self-interest, and by social
interactions, which push the member to seek consensus within the group. When
$D_{kj}=0$, for $j=1,2,...N$ the $k$-th member possesses no knowledge about
the fitness function, and his choices are driven only by consensus seeking.
Note that the choice configuration that optimizes the perceived fitness Eq.
(\ref{perceived fitness}), does not necessarily optimize the group fitness Eq.
(\ref{fitness function}). This makes the mechanism of social interactions, by
means of which knowledge is transferred, crucial for achieving high-performing
decision-making process. We build the matrix $\mathbf{D}$, by randomly
choosing $D_{kj}=1$ with probability $p\in\left[  0,1\right]  $, and
$D_{kj}=0$ with probability $1-p$. By increasing $p$ from $0$ to $1$ we
control the level of knowledge of the members, which affects the ability of
the group in maximizing the fitness function Eq. (\ref{fitness function}).

All members of the group make choices on each of the $N$ decision variables
$d_{j}$. Therefore, the state of the $k$-th member ($k=1,2,..,M$) is
identified by the $N$-dimensional vector $\mathbf{\sigma}_{k}=\left(
\sigma_{k}^{1},\sigma_{k}^{2},...\sigma_{k}^{N}\right)  $, where $\sigma
_{k}^{j}=\pm1$ is a binary variable representing the opinion of the $k$-th
member on the $j$-th decision. For any given decision variable $d_{j}$,
individuals $k$ and $h$ agree if $\sigma_{k}^{j}=\sigma_{h}^{j}$, otherwise
they disagree. Within the framework of Ising's approach \cite{IsingOriginal,
IsingOr2, IS1}, disagreement is characterized by a certain level of conflict
$E_{kh}^{j}$ (energy level) between the two socially interacting members $k$
and $h$, i.e. $E_{kh}^{j}=-J\sigma_{k}^{j}\sigma_{h}^{j}$, where $J$ is the
strength of the social interaction. Therefore, the total level of conflict on
the decision $d_{j}$ is given by:
\begin{equation}
E^{j}=-\sum_{\left(  k,h\right)  }J\sigma_{k}^{j}\sigma_{h}^{j}
\label{Level of conflict layer}%
\end{equation}
where the symbol $\left(  {}\right)  $ indicates that the sum is limited to
the nearest neighbors, i.e. to those individuals which are directly connected
by a social link.\begin{figure}[ptb]
\begin{center}
\includegraphics[
height=3.56cm,
width=8.0cm
]{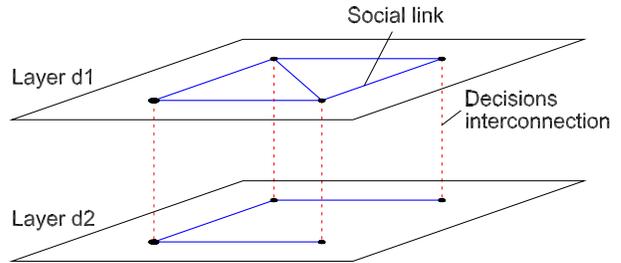}
\end{center}
\caption{A two-layer multiplex network. Each layer is associated with one
single decision variable. Blue lines represent social links between members,
whereas dashed red lines represent the interconnections between the decisions
of each member.}%
\label{Figure1}%
\end{figure}

A multiplex network \cite{Multeplx1, Intro-MutiPlNet, multi3, multi4, multi5}
with $N$ different layers is defined. On each layer, individuals share their
opinions on a certain decision variable $d_{j}$ leading to a certain level of
conflict $E^{j}$. The graph of social network on the layer $d_{j}$ is
described in terms of the symmetric adjacency matrix $\mathbf{A}^{j}$ with
elements $A_{kh}^{j}$. The interconnections between different layers represent
the interactions among the opinions of the same individual $k$ on the decision
variables. Figure \ref{Figure1} shows an example of a multiplex network with
only two layers, where the dashed lines connecting the different decision
layers represent the interaction between the opinions that each member has on
the decision variables. This interaction occurs via the $N-K$ perceived
fitness, i.e. changing the opinion on the decision variable $j$ causes a
modification of the perceived pay-off, which also depends on the opinions the
member has on the remaining decision variables. In order to model the dynamics
of decision-making in terms of a continuous-time Markov process, we define the
state vector $\mathbf{s}$ of the entire group as $\mathbf{s=}\left(
s_{1},s_{2},...,s_{n}\right)  =\left(  \sigma_{1}^{1},\sigma_{1}^{2}%
,...\sigma_{1}^{N},\sigma_{2}^{1},\sigma_{2}^{2},...\sigma_{2}^{N}%
,...,\sigma_{M}^{1},\sigma_{M}^{2},...\sigma_{M}^{N}\right)  $ of size $n$
$=M\times N$, and the block diagonal adjacency matrix $\mathbf{A}%
=\mathrm{diag}\left(  \mathbf{A}^{1},\mathbf{A}^{2},...,\mathbf{A}^{N}\right)
$. For any given $l$-th component $s_{l}=\sigma_{k}^{j}$ of the vector
$\mathbf{s}$ it is possible to uniquely identify the member $k$ and the
decision variable $j$ by means of the relations $k=\mathrm{quotient}\left(
l-1,M\right)  +1$, and $j=\operatorname{mod}\left(  l-1,M\right)  +1$. The
total level of conflict can be then rephrased as%
\begin{equation}
E\left(  \mathbf{s}\right)  =-\frac{1}{2}J\mathbf{As\cdot s=}-\frac{1}{2}%
J\sum_{ij}A_{ij}s_{i}s_{j}\label{total energy level}%
\end{equation}
Observe that $A_{ii}=0$ (with $i=1,...,n$). In Eq. (\ref{total energy level})
the term $1/2$ avoids that each couple of members $k$ and $h$ be double
counted. Now let be $P\left(  \mathbf{s},t\right)  $ the probability that, at
time $t$, the state vector takes the value $\mathbf{s}$ out of $2^{n}$
possible states. The time evolution of the probability $P\left(
\mathbf{s},t\right)  $ obeys the master equation%
\begin{align}
\frac{dP\left(  \mathbf{s},t\right)  }{dt} &  =-\sum_{l}w\left(
\mathbf{s}_{l}\rightarrow\mathbf{s}_{l}^{\prime}\right)  P\left(
\mathbf{s}_{l},t\right)  \label{Markov chain}\\
&  +\sum_{l}w\left(  \mathbf{s}_{l}^{\prime}\rightarrow\mathbf{s}_{l}\right)
P\left(  \mathbf{s}_{l}^{\prime},t\right)  \nonumber
\end{align}
where $\mathbf{s}_{l}=\left(  s_{1},s_{2},...,s_{l},...,s_{n}\right)  $,
$\mathbf{s}_{l}^{\prime}=\left(  s_{1},s_{2},...,-s_{l},...,s_{n}\right)  $.
The transition rate $w\left(  \mathbf{s}_{l}\rightarrow\mathbf{s}_{l}^{\prime
}\right)  $ is the probability per unit time that the opinion $s_{l}$ flips to
$-s_{l}$ while the others remain temporarily fixed. Recalling that flipping of
opinions is governed by social interactions and self-interest a possible
ansatz for the transition rates is%
\begin{align}
w\left(  \mathbf{s}_{l}\rightarrow\mathbf{s}_{l}^{\prime}\right)   &
=\frac{1}{2}\left[  1-s_{l}\tanh\left(  \beta J\sum_{h}A_{lh}s_{h}\right)
\right]  \label{transition rates.}\\
&  \times\exp\left\{  \beta^{\prime}\left[  \Delta V\left(  \mathbf{s}%
_{l}^{\prime},\mathbf{s}_{l}\right)  \right]  \right\}  \nonumber
\end{align}
In Eq. (\ref{transition rates.}) the pay-off function $\Delta V\left(
\mathbf{s}_{l}^{\prime},\mathbf{s}_{l}\right)  =\bar{V}\left(  \mathbf{s}%
_{l}^{\prime}\right)  -\bar{V}\left(  \mathbf{s}_{l}\right)  $, where $\bar
{V}\left(  \mathbf{s}_{l}\right)  =V_{k}\left(  \mathbf{\sigma}_{k}\right)  $,
is simply the change of the fitness perceived by the agent
$k=\mathrm{quotient}\left(  l-1,M\right)  +1$, when its opinion $s_{l}%
=\sigma_{k}^{j}$ on the decision $j=\mathrm{\operatorname{mod}}\left(
l-1,M\right)  +1$ changes from $s_{l}=\sigma_{k}^{j}$ to $s_{l}^{\prime
}=-\sigma_{k}^{j}$. The transition rates $w\left(  \mathbf{s}_{l}%
\rightarrow\mathbf{s}_{l}^{\prime}\right)  $ have been chosen to be the
product of the transition rate of the Ising-Glauber dynamics \cite{Glauber}
(see also Appendix \ref{Glauber-general}), and the Weidlich exponential rate
$\exp\left\{  \beta^{\prime}\left[  \Delta V\left(  \mathbf{s}_{k}^{\prime
},\mathbf{s}_{k}\right)  \right]  \right\}  $\cite{weidlich2, Sweitzer}. Note
that Eq. (\ref{transition rates.}) satisfies the detailed balance condition
(see Appendix \ref{balance cond}). In Eq. (\ref{transition rates.}) the
quantity $\beta$ is the inverse of the so-called social temperature and is a
measure of the chaotic circumstances, which lead to a random opinion change.
The term $\beta^{\prime}$ is related to the degree of uncertainty associated
with the information about the perceived fitness (the higher $\beta^{\prime}$
the less the uncertainty).

To solve the Markov process Eqs. (\ref{Markov chain}, \ref{transition rates.}%
), we employ a simplified version of the exact stochastic simulation algorithm
proposed by Gillespie \cite{Gillespie1, Gillespie2}. A brief summary of the
algorithm is provided in Appendix \ref{Gillespie_alg}. The algorithm allows to
generate a statistically correct trajectory of the stochastic process Eqs.
(\ref{Markov chain}, \ref{transition rates.}).

\section{Measuring the performance of the collective decision-making process}

The group fitness value Eq. (\ref{fitness function}) and the level of
agreement between the members (i.e. social consensus) are used to measure the
performance of the collective-decision making process. To calculate the group
fitness value, the vector $\mathbf{d=}\left(  d_{1},d_{2},...,d_{N}\right)  $
needs to be determined. To this end, consider the set of opinions $\left(
\sigma_{1}^{j},\sigma_{2}^{j},...,\sigma_{M}^{j}\right)  $ that the members of
the group have about the decision $j$, at time $t$. The decision $d_{j}$ is
obtained by employing the majority rule, i.e. we set
\begin{equation}
d_{j}=\mathrm{sgn}\left(  M^{-1}\sum_{k}\sigma_{k}^{j}\right)  ,\qquad
j=1,2,...,N \label{majority rule}%
\end{equation}
If $M$ is even and in the case of a parity condition, $d_{j}$ is, instead,
uniformly chosen at random between the two possible values $\pm1$. The group
fitness is then calculated as $V\left[  \mathbf{d}\left(  t\right)  \right]  $
and the ensemble average $\left\langle V\left(  t\right)  \right\rangle $ is
then evaluated. The efficacy of the group in optimizing $\left\langle V\left(
t\right)  \right\rangle $ is then calculated in terms of normalized average
fitness $\left\langle V\left(  t\right)  \right\rangle /V_{\max}$ where
$V_{\max}=\max\left[  V\left(  \mathbf{d}\right)  \right]  $.

The consensus of the members on the decision variable $j$ is measured as
follows. We define the average opinion $\bar{\sigma}^{j}$ of the group on the
decision $j$%
\begin{equation}
\bar{\sigma}^{j}=\frac{1}{M}\sum_{k}\sigma_{k}^{j}
\label{average opinion on j}%
\end{equation}
Note that the quantity $\bar{\sigma}^{j}$ ranges in the interval $-1\leq
\bar{\sigma}^{j}\leq1$, and that $\bar{\sigma}^{j}=\pm1$ only when full
consensus is reached. Therefore, a possible measure of the consensus among the
members on the decision variable $j$ is given by the ensemble average of the
time-dependent quantity $C^{j}=\left(  \bar{\sigma}^{j}\right)  ^{2}\in\left[
0,1\right]  $, i.e.,
\begin{equation}
\left\langle C^{j}\left(  t\right)  \right\rangle =\frac{1}{M^{2}}\sum
_{kh}\left\langle \sigma_{k}^{j}\left(  t\right)  \sigma_{h}^{j}\left(
t\right)  \right\rangle =\frac{1}{M^{2}}\sum_{kh}R_{hk}^{j}\left(  t\right)
\label{consensus measure}%
\end{equation}
Note that $\left\langle \sigma_{k}^{j}\left(  t\right)  \sigma_{h}^{j}\left(
t\right)  \right\rangle =R_{hk}^{j}\left(  t\right)  $ is the correlation
function of the opinions of the members $k$ and $h$ on the same decision
variable $j$. Given this, a possible ansatz to measure the entire consensus of
the group on the whole set of decisions is
\begin{equation}
\left\langle C\left(  t\right)  \right\rangle =\frac{1}{N}\sum_{j}\left\langle
C^{j}\left(  t\right)  \right\rangle =\frac{1}{M^{2}N}\sum_{j=1}^{N}%
\sum_{kh=1}^{M}R_{hk}^{j}\left(  t\right)  \label{whole consensus}%
\end{equation}
Note that $0\leq\left\langle C\left(  t\right)  \right\rangle \leq1$.

\section{Simulation and results}

We consider, unless differently specified, a group of $M=6$ members which have
to make $N=12$ decisions. For the sake of simplicity, the network of social
interactions on each decision layer $j$ is described by a complete graph,
where each member is connected to all the others. We also set $\beta^{\prime
}=10,$ since we assume that the information about the perceived fitness
function is characterized by a low level of uncertainty. We simulate many
diverse scenarios to investigate the influence of the parameter $p$, i.e. of
the level of knowledge of the members, and the effect of the parameter $\beta
J$ on the final outcome of the decision-making process. The simulation is
stopped at steady-state. This condition is identified by simply taking the
time-average of consensus and pay-off over consecutive time intervals of fixed
length $T\ $and by checking that the difference between two consecutive
averages is sufficiently small. \begin{figure}[ptb]
\begin{center}
\includegraphics[
height=16.38cm,
width=8.0cm
]{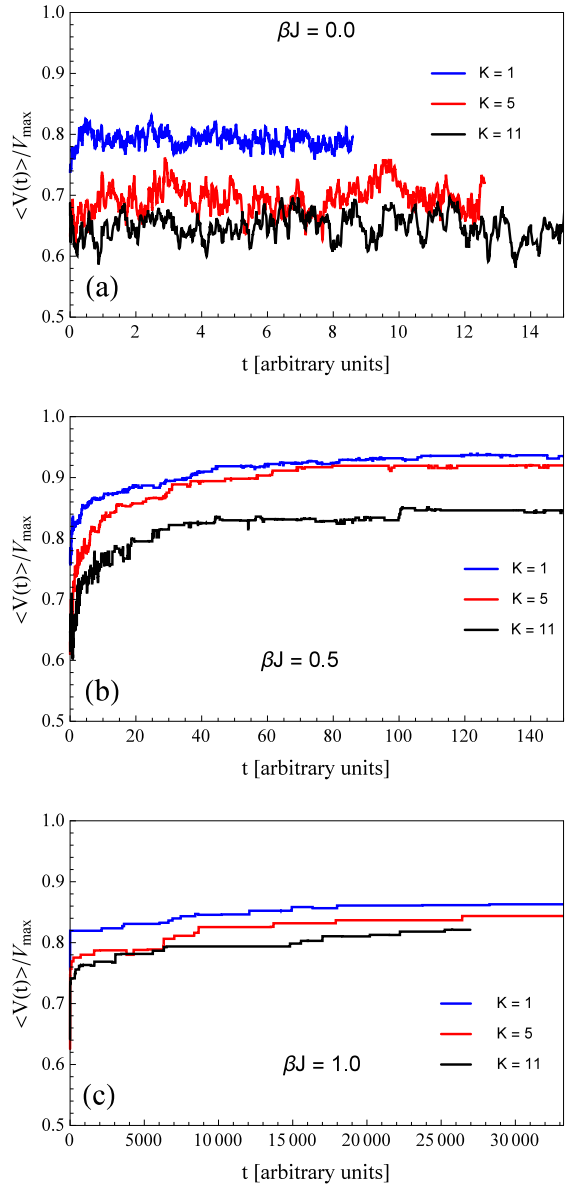}
\end{center}
\caption{The time-evolution of the normalized average group fitness
$\left\langle V\left(  t\right)  \right\rangle /V_{\max}$, for $p=0.5$,
$K=1,5,11$. $\beta J=0.0$, (a); $\beta J=0.5$, (b); $\beta J=1.0$, (c).}%
\label{Figure2}%
\end{figure}For any given $p$ and $\beta J$, each stochastic process Eqs.
(\ref{Markov chain}, \ref{transition rates.}) is simulated by generating 100
different realizations (trajectories). For each single realization, the
competence matrix $\mathbf{D}$ is set, and the initial state of the system is
obtained by drawing from a uniform probability distribution, afterwards the
time evolution of the state vector is calculated with the stochastic
simulation algorithm (see Appendix \ref{Gillespie_alg}). \begin{figure}[ptb]
\begin{center}
\includegraphics[
height=16.38cm,
width=8.0cm
]{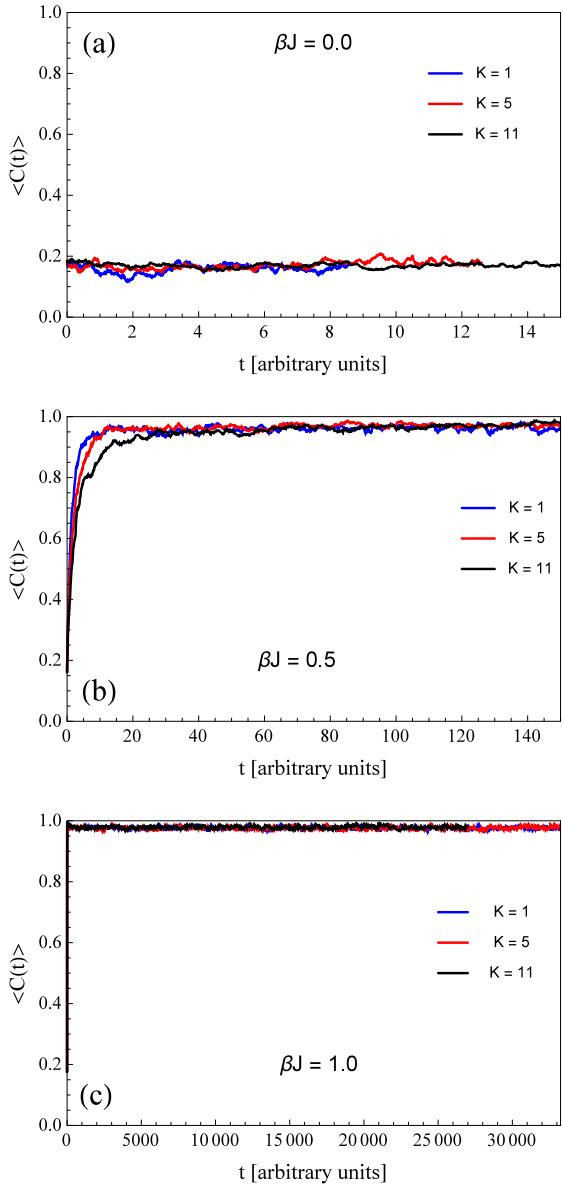}
\end{center}
\caption{The time-evolution of the statistically averaged consensus
$\left\langle C\left(  t\right)  \right\rangle $ for $p=0.5$, $K=1,5,11$.
$\beta J=0.0$, (a); $\beta J=0.5$, (b); $\beta J=1.0$, (c).}%
\label{Figure3}%
\end{figure}Fig. \ref{Figure2} shows the time-evolution of normalized average
fitness $\left\langle V\left(  t\right)  \right\rangle /V_{\max}$, for $p=0.5$
(i.e. for a moderate level of knowledge of the members), different values of
the complexity parameter $K=1,5,11$, and different values of $\beta
J=0.0,0.5,1.0$. We observe that for $\beta J=0$, i.e. in absence of social
interactions [see Fig. \ref{Figure2}(a)] the decision-making process is
strongly inefficient, as witnessed by the very low value of the average
fitness of the group. Each individual of the group makes his/her choices in
order to optimize the perceived fitness, but, because of the absence of social
interactions, he/she behaves independently from the others and does not
receive any feedback about the actions of the other group members. Hence,
individuals remain close to their local optima, group fitness cannot be
optimized [see Fig. \ref{Figure2}(a)], and the consensus is low [see Fig.
\ref{Figure3}(a)].\begin{figure}[ptb]
\begin{center}
\includegraphics[
height=10.71cm,
width=8.0cm
]{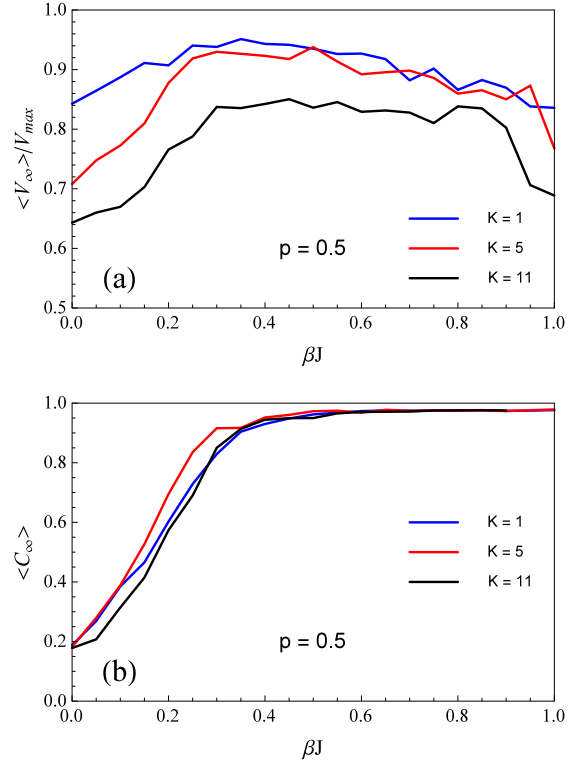}
\end{center}
\caption{The stationary values of the normalized averaged fitness
$\left\langle V_{\infty}\right\rangle /V_{\max}$ as a function of $\beta J$,
(a); and of the statistically averaged consensus $\left\langle C_{\infty
}\right\rangle $ as a function of $\beta J$, (b). Results are presented for
$p=0.5$, $K=1,5,11$}%
\label{Figure4}%
\end{figure}\begin{figure}[ptb]
\begin{center}
\includegraphics[
height=10.93cm,
width=8.0cm
]{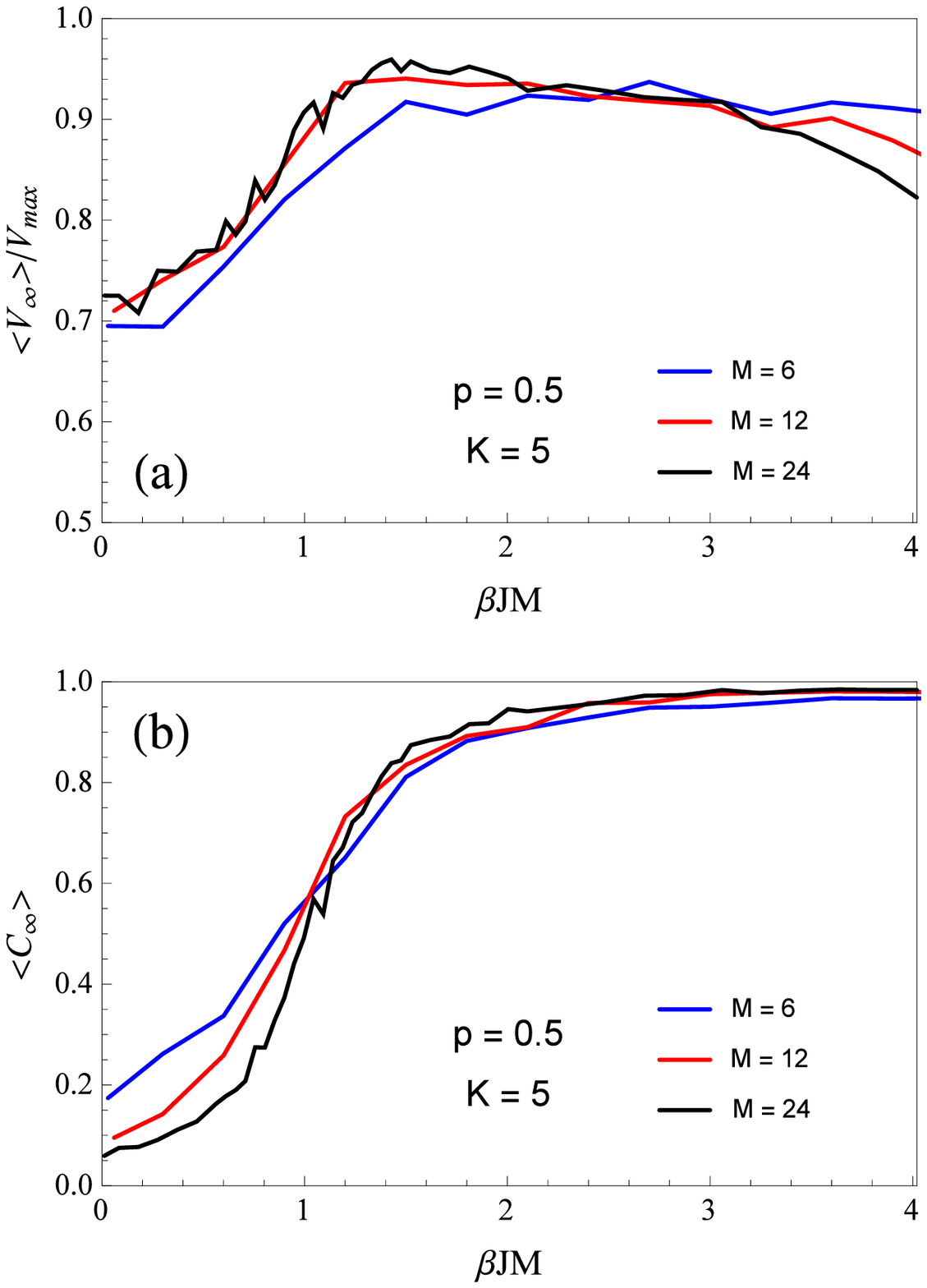}
\end{center}
\caption{The stationary values of the normalized averaged fitness
$\left\langle V_{\infty}\right\rangle /V_{\max}$ as a function of $M\beta J$,
(a); and of the statistically averaged consensus $\left\langle C_{\infty
}\right\rangle $ as a function of $M\beta J$, (b). Results are presented for
$p=0.5$, $K=5$ and for three different team sizes: $M=6,12,24$.}%
\label{Figure-criticality}%
\end{figure}As the strength of social interactions increases, i.e., $\beta
J=0.5$ [Fig. \ref{Figure2}(b)], members can exchange information about their
choices. Social interactions push the individuals to seek consensus with the
member who is experiencing higher payoff. In fact, on the average, those
members, which find a higher increase of their perceived fitness, change
opinion much faster than the others. Thus, the other members, in process of
seeking consensus, skip the local optima of their perceived fitness and keep
exploring the landscape, leading to a substantial increase of the group
performance both in terms of group fitness values [Fig. \ref{Figure2}(b)] as
well as in terms of final consensus [Fig. \ref{Figure3}(b)].
\begin{figure}[ptb]
\begin{center}
\includegraphics[
height=16.38cm,
width=8.0cm
]{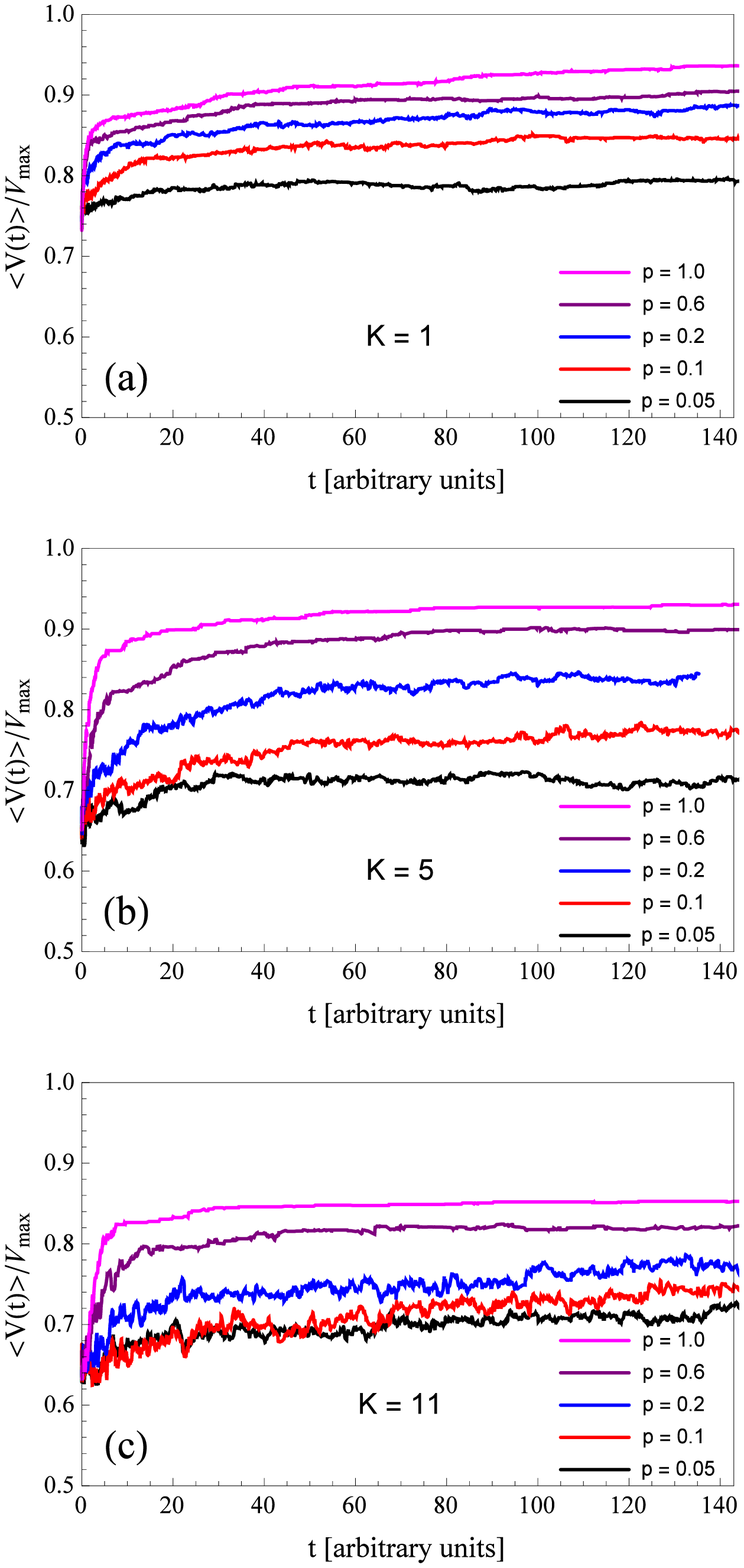}
\end{center}
\caption{The time-evolution of the normalized average group fitness
$\left\langle V\left(  t\right)  \right\rangle /V_{\max}$ for $\beta J=0.5$,
$p=0.05,0.1,0.2,0.6,1$. $K=1$, (a); $\ K=5$, (b); $K=11$, (c).}%
\label{Figure5}%
\end{figure}Thus, the system collectively shows a higher level of knowledge
and higher ability in making good choices than the single members (i.e., a
higher degree of intelligence). It is noteworthy that when the strength of
social interactions is too large, $\beta J=1$, [Fig. \ref{Figure2}(c)] the
performance of the group in terms of fitness value worsens. In fact, very high
values of $\beta J$, accelerating the achievement of consensus among the
members [Fig. \ref{Figure3}(c)], significantly impede the exploration of the
fitness landscape and hamper that change of opinions can be guided by payoff
improvements. \begin{figure}[ptb]
\begin{center}
\includegraphics[
height=16.38cm,
width=8.0cm
]{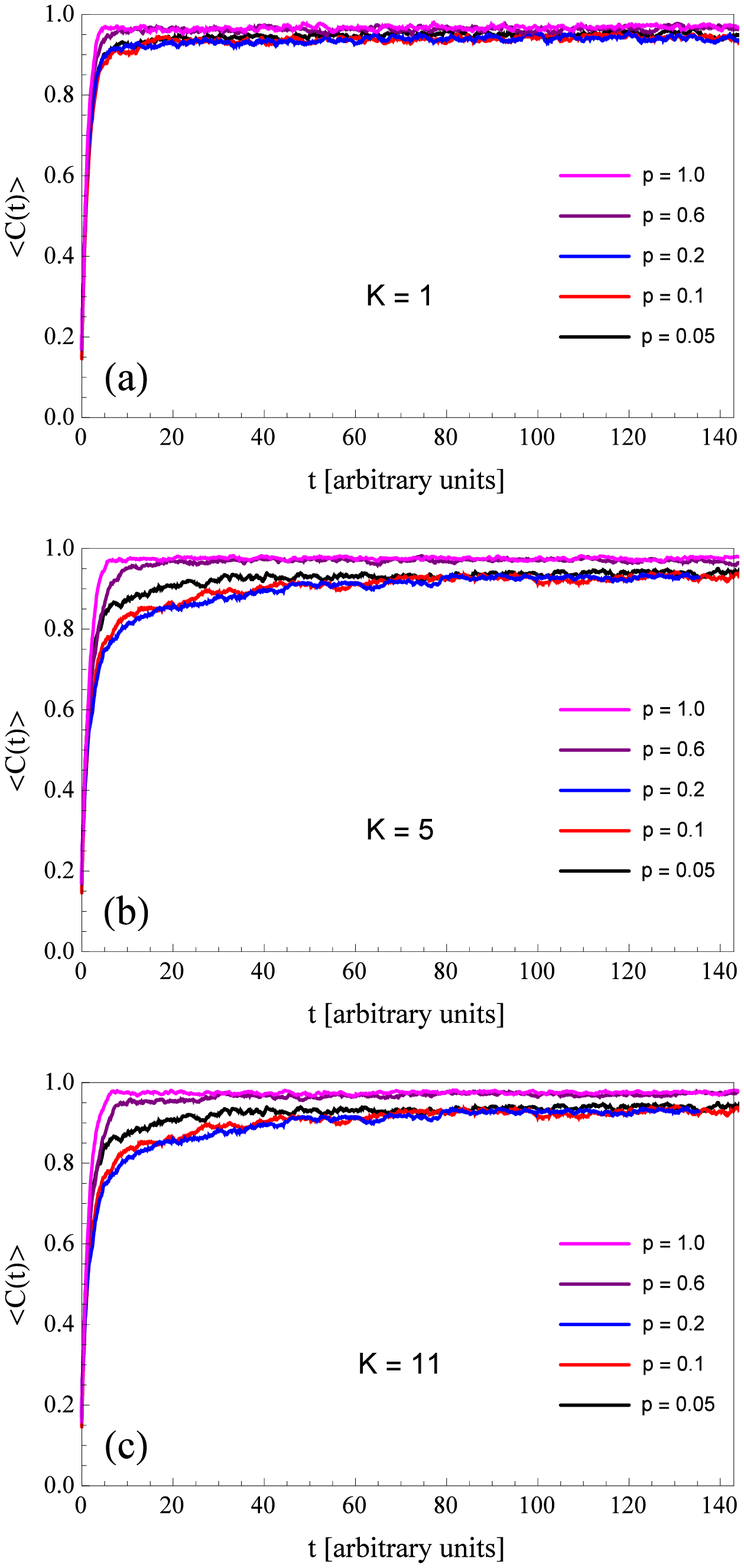}
\end{center}
\caption{The time-evolution of the statistically averaged consensus
$\left\langle C\left(  t\right)  \right\rangle $ for $\beta J=0.5$,
$p=0.05,0.1,0.2,0.6,1$. $K=1$, (a); $\ K=5$, (b); $K=11$, (c).}%
\label{Figure6}%
\end{figure}The search of the optimum on the fitness landscape is slowed down,
and the performance of the collective decision-making decreases both in terms
of the time required to reach the steady-state as well as in terms of group fitness.

Figure \ref{Figure2} shows that rising the complexity of the landscape, i.e.
increasing $K$, negatively affects the performance of the collective
decision-making process, but does not qualitatively change the behavior of the
system. However, Figure \ref{Figure2}(b) also shows that, in order to cause a
significant worsening of the group fitness, $K$ must take very large values,
i.e., $K=11$. Instead, at moderate, but still significant, values of
complexity (see results for $K=5$) the decision-making process is still very
effective, leading to final group fitness values comparable to those obtained
at the lowest level of complexity, i.e., at $K=1$.

In Figure \ref{Figure3} the ensemble average $\left\langle C\left(  t\right)
\right\rangle $ of the consensus among the members is shown as a function of
time $t$, for $p=0.5$, $K=1,5,11$, and for different values of $\beta
J=0.0,0.5,1.0$. At $\beta J=0$, the consensus is low. \begin{figure}[ptb]
\begin{center}
\includegraphics[
height=10.71cm,
width=8.0cm
]{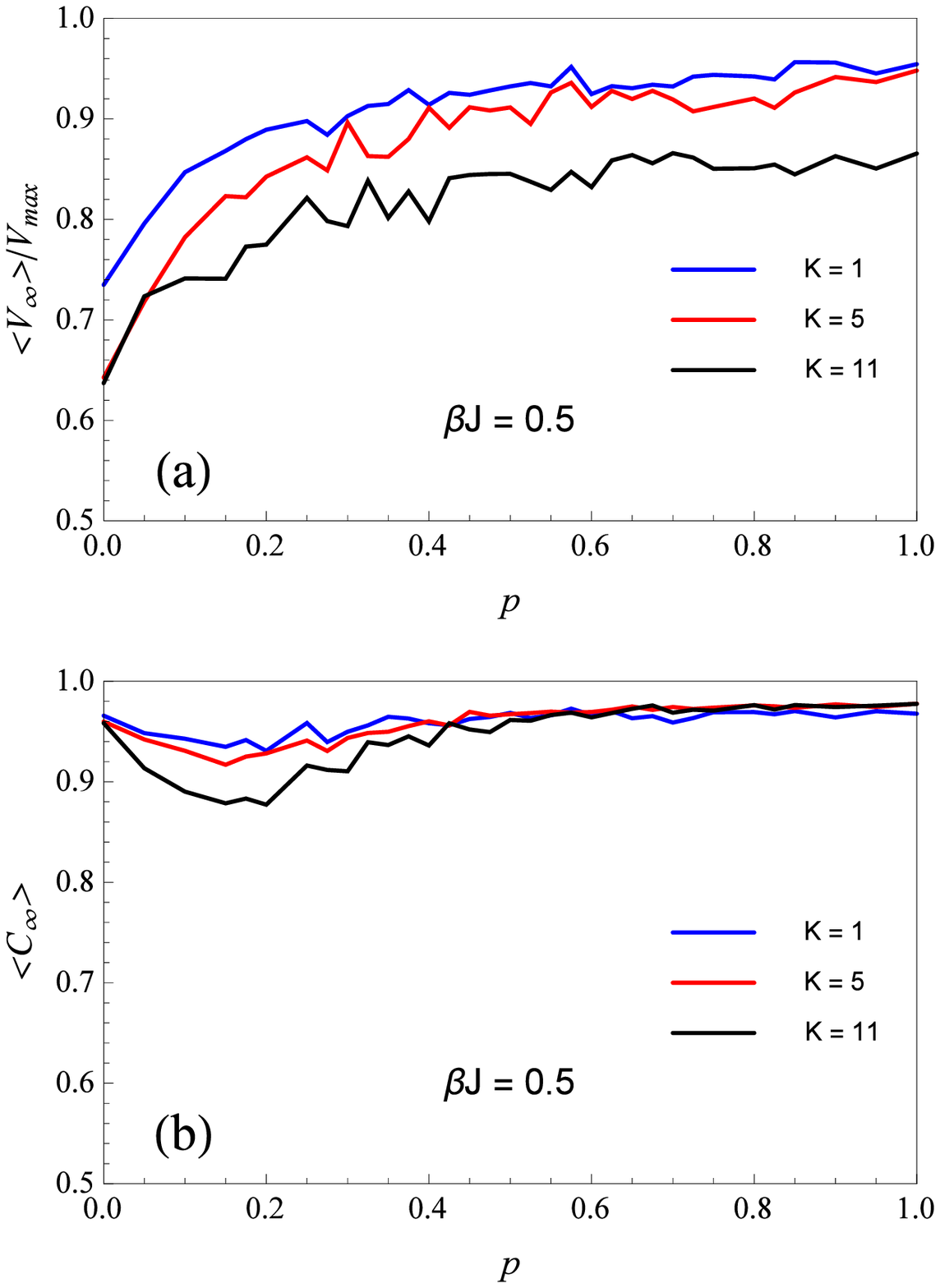}
\end{center}
\caption{The stationary values of the normalized average group fitness
$\left\langle V_{\infty}\right\rangle /V_{\max}$ as a function of $p$, (a);
and of the statistically averaged consensus $\left\langle C_{\infty
}\right\rangle $ as a function of $p$, (b). Results are presented for $\beta
J=0.5$, $K=1,5,11$}%
\label{Figure7}%
\end{figure}In this case, at each time $t$, members' opinions are random
variables almost uniformly distributed between the two states $\pm1$. Hence,
the quantity $\left\langle C\left(  t\right)  \right\rangle $ can be
analytically calculated as $\left\langle C\left(  t\right)  \right\rangle
\approx1/M$. For $M=6$ this gives $\left\langle C\left(  t\right)
\right\rangle \approx0.16$, which is just the average value observed in Fig.
\ref{Figure3}(a). As the strength of social interactions rises, members more
easily converge toward a common opinion. However, the random nature of the
opinion dynamics still prevents full agreement from being achieved, see Fig.
\ref{Figure3}(b). This, as observed in Figure \ref{Figure2}(b), has a very
beneficial effect as individuals continue exploring the fitness landscape
looking for maxima, thus leading to higher performance of the collective
decision-making process. However, when the strength of social interactions is
significantly increased, a very high value of consensus among members is
rapidly achieved [see Fig. \ref{Figure3}(c)], the exploration of the landscape
is slowed down, and the performance of the decision making-process
significantly worsen, see Figure \ref{Figure2}(c). We, then, expect that,
given $\beta^{\prime}$and $K$, a optimum of $\beta J$ exists, which maximizes
the steady-state fitness of the group. This is, indeed, confirmed by the
analysis shown in Figure \ref{Figure4}, where the steady-state values of the
normalized group fitness $\left\langle V_{\infty}\right\rangle /V_{\max
}=\left\langle V\left(  t\rightarrow\infty\right)  \right\rangle /V_{\max}$
[Fig. \ref{Figure4}(a)], and social consensus $\left\langle C_{\infty
}\right\rangle =\left\langle C\left(  t\rightarrow\infty\right)  \right\rangle
$ [Fig. \ref{Figure4}(b)] are plotted as a function of $\beta J$, for $p=0.5$
and the three considered values of $K=1,5,11$. results in Figure
\ref{Figure4}(a) stresses that the fitness landscape complexity (i.e., the
parameter $K$) marginally affects the performance of the decision-making
process in terms of group fitness, provided that $K$ does not take too high
values. In fact curves calculated for $K=1,5$ run close to each-other. More
interesting, Figure \ref{Figure4} shows that increasing $\beta J$ from zero,
makes both $\left\langle V_{\infty}\right\rangle /V_{\max}$ and $\left\langle
C_{\infty}\right\rangle $ rapidly increase. This increment is, then, followed
by a region of a slow change of $\left\langle V_{\infty}\right\rangle
/V_{\max}$ and $\left\langle C_{\infty}\right\rangle $. It is worth noticing,
that the highest group fitness value is obtained at the boundary between the
increasing and almost stationary regions of $\left\langle C_{\infty
}\right\rangle $. Moreover, results show that high consensus is necessary to
guarantee high efficacy of the decision-making process, i.e. high values of
$\left\langle V_{\infty}\right\rangle /V_{\max}$. This suggests that the
decision making becomes optimal, i.e. the group as a whole is characterized by
a higher degree of intelligence, at the point where the system dynamics
changes qualitatively. This aspect of the problem is investigated in Figure
\ref{Figure-criticality} where the stationary values of normalized
group-fitness and consensus are shown as a function of the quantity $\beta JM$
for different sizes $M=6,12,24$, for $N=12$ and $K=5$. Notably, the transition
from low to high fitness values is always accompanied by an analogous
transition from low to high consensus of the group. The transition becomes
sharper and sharper as the group size $M$ is incremented. In all cases the
transition occurs for $\beta JM\approx1$. This value is particularly
interesting as it can be easily shown, by using a mean-field approach (see
Appendix \ref{mean-field}), that for large Ising systems $M\gg1$ and in the
case of complete graphs, the critical values $\left(  \beta J\right)  _{c}$,
at which consensus sets in, satisfies the relation $\left(  \beta J\right)
_{c}=1/M$, in perfect agreement with our small-size numerical calculations.
This is a very important result, which has analogies in many different
self-organizing systems, as flocking systems and information flow processing
\cite{Flocking 1, Flocking2, Info1,Info2, Info3}.

In Figure \ref{Figure5} we investigate the influence of the level of knowledge
$p$ of members on the time-evolution of the normalized average fitness
$\left\langle V\left(  t\right)  \right\rangle /V_{\max}$. Results are
presented for $\beta J=0.5$, $K=1,5,11$, and for different values of
$p=0.05,0.1,0.2,0.6,1.0$. Results show that improving the knowledge of the
members, i.e. increasing $p$, enhances the performance of the decision-making
process. In particular, a higher steady-state normalized fitness $\left\langle
V_{\infty}\right\rangle /V_{\max}$, and a faster convergence toward the
steady-state are observed. Note also, that, especially in the case of high
complexity [Figure \ref{Figure5}(b,c)], increasing $p$ above $0.2$ reduces the
fluctuations of $\left\langle V\left(  t\right)  \right\rangle $, as a
consequence of the higher agreement achieved among the members at higher level
of knowledge. This is clear in Figure \ref{Figure6}, where the time-evolution
of the consensus $\left\langle C\left(  t\right)  \right\rangle $ is shown for
$\beta J=0.5$, $K=1,5,11$, and $p=0.05,0.1,0.2,0.6,1.0$. In Figure
\ref{Figure7} the steady-state values of the normalized group fitness
$\left\langle V_{\infty}\right\rangle /V_{\max}$ [Fig. \ref{Figure7}(a)], and
social consensus $\left\langle C_{\infty}\right\rangle $ [Fig. \ref{Figure7}%
(b)] are shown as a function of $p$, for $\beta J=0.5$ and the three
considered values of $K=1,5,11$. Note that as $p$ is increased from zero, the
steady state value $\left\langle V_{\infty}\right\rangle /V_{\max}$ initially
grows fast [Fig. \ref{Figure7}(a)]. In fact, because of social interactions,
increasing the knowledge of each member also increases the knowledge of the
group as a whole. But, above a certain threshold of $p$ the increase of
$\left\langle V_{\infty}\right\rangle /V_{\max}$ is much less significant.
This indicates that the knowledge of the group is subjected to a saturation
effect. Therefore, a moderate level of knowledge is already enough to
guarantee very good performance of decision-making process, higher knowledge
levels being only needed to accelerate the convergence of the decision-making
process. Figure \ref{Figure7}(b) shows that for vanishing values of $p$ the
consensus $\left\langle C_{\infty}\right\rangle $ takes high values, as each
member's choice is driven only by consensus seeking. Increasing $p$ initially
causes a decrease of consensus, as the self-interest of each member leads to a
certain level of disagreement. However, a further increment of $p$ makes the
members' knowledge overlap so that the self-interest of each member almost
points in the same direction, resulting in a consensus increase.

\section{Conclusions}

In this paper we developed a model of collective decision-making in human
groups in presence of complex environment. The model described the time
evolution of group choices in terms of a time-continuous Markov process, where
the transition rates have been defined so as to capture the effect of the two
main forces, which drive the change of opinion of the members of the group.
These forces are the rational behavior which pushes each member to increase
his/her self-interest, and the social interactions, which push the members to
reach a common opinion. Our study provides contribution to the literature
identifying under which circumstances collective decision making is more
performing. We found that a moderate strength of social interactions allows
for knowledge transfer among the members, leading to higher knowledge level of
the group as a whole. This mechanism, coupled with the ability to explore the
fitness landscape, strongly improves the performance of the decision-making
process. In particular we found that the threshold value of the social
interaction strength, at which the entire group behaves as unique entity
characterized by a higher degree of intelligence, is just the critical
threshold at which the consensus among the members sets in. This value can be
also calculated for large systems trough mean-field techniques and results to
be in perfect agreement with our small system calculations. One can therefore
estimate for any given social temperature $\beta^{-1}$ and social interaction
strength $J$ the optimal number $M_{\mathrm{opt}}\approx\left(  \beta
J\right)  ^{-1}$ of team members leading to the emergence of a superior
intelligence of the group.

We also found that increasing the level of knowledge of the members improves
performance. However, above a certain threshold the knowledge of the group
saturates, i.e. the performance of the collective decision-making process
becomes much less sensitive to the level of knowledge of each single member.
Therefore, we can state that the collective decision-making is very
high-performing already at moderate level of knowledge of the members, and
that very high knowledge of all members only serves to accelerate the
convergence of the decision-making process. Our results also showed that human
groups with optimal levels of members' knowledge and strength of social
interactions very well manage complex problems.

\appendix

\section{The $N-K$ fitness landscape generation\label{NK-model}}

In the $N-K$ model a real valued fitness is assigned to each bit string
$\mathbf{d=}\left(  d_{1},d_{2},...,d_{N}\right)  $, where $d_{i}=\pm1$. This
is done by first assigning a real valued contribution $W_{i}$ to the $i$-th
bit $d_{i}$, and then by defining the fitness function as $V\left(
\mathbf{d}\right)  =N^{-1}\sum_{j=1}^{N}W_{j}\left(  d_{j},d_{1}^{j},d_{2}%
^{j},..,d_{K}^{j}\right)  $. Each contribution $W_{i}$ depends not just on $i$
and $d_{i}$ but also on $K$ ($0\leq K<N$) other bits. Now let us define the
substring $\mathbf{s}_{i}=\left(  d_{i},d_{1}^{i},d_{2}^{i},..,d_{K}%
^{i}\right)  $, by chosing at random, for each bit $i$, $K$ other bits. Each
single contribution $W_{i}\left(  \mathbf{s}_{i}\right)  $ is then a random
function of $2^{K+1}$ possible values of $\mathbf{s}_{i}$, and its value is
drawn from a uniform distribution. Thus, a random table\ of contributions is
generated independently for each $i$-th bit, thus allowing the calculate of
the fitness function $V\left(  \mathbf{d}\right)  $. The reader is referred to
Refs. \cite{NK-model, NK-model1, NK-model2} for more details on the $N-K$
complex landscapes.

\section{The Glauber dynamics on general graphs\label{Glauber-general}}

Consider the Ising model on a general graph with adjacency matrix $A_{ij}$.
The total energy of the system is given in Eq. (\ref{total energy level}). In
steady state conditions the stationary distribution of the probability of the
states $P_{0}\left(  \mathbf{s}\right)  $ is given by the Boltzmann
distribution%
\begin{equation}
P_{0}\left(  \mathbf{s}\right)  =\frac{\exp\left[  -\beta E\left(
\mathbf{s}\right)  \right]  }{Z} \label{boltzmann distrib}%
\end{equation}
where $Z=\sum_{l}\exp\left[  -\beta E\left(  \mathbf{s}_{l}\right)  \right]  $
is the partition function of the system. The detailed balance condition then
requires that%
\begin{equation}
\frac{w\left(  \mathbf{s}_{l}\rightarrow\mathbf{s}_{l}^{\prime}\right)
}{w\left(  \mathbf{s}_{l}^{\prime}\rightarrow\mathbf{s}_{l}\right)  }%
=\frac{P_{0}\left(  \mathbf{s}_{l}^{\prime}\right)  }{P_{0}\left(
\mathbf{s}_{l}\right)  }=\frac{\exp\left[  -\beta E\left(  \mathbf{s}%
_{l}^{\prime}\right)  \right]  }{\exp\left[  -\beta E\left(  \mathbf{s}%
_{l}\right)  \right]  } \label{detailed balance gluaber}%
\end{equation}
Now observe that
\begin{equation}
E\left(  \mathbf{s}_{l}\right)  =-Js_{l}\sum_{j}A_{lj}s_{j}-\frac{1}{2}%
J\sum_{ij\neq l}A_{ij}s_{i}s_{j} \label{energy1}%
\end{equation}
Substituting Eqs. (\ref{energy1}) in Eq. (\ref{detailed balance gluaber}) and
recalling that $\exp\left(  x\right)  =\cosh\left(  x\right)  +\sinh\left(
x\right)  $ we get%
\begin{equation}
\frac{w\left(  \mathbf{s}_{l}\rightarrow\mathbf{s}_{l}^{\prime}\right)
}{w\left(  \mathbf{s}_{l}^{\prime}\rightarrow\mathbf{s}_{l}\right)  }%
=\frac{1-\tanh\left(  \beta Js_{l}\sum_{j}A_{lj}s_{j}\right)  }{1+\tanh\left(
\beta Js_{l}\sum_{j}A_{lj}s_{j}\right)  } \label{primo}%
\end{equation}
Noting that $s_{l}=\pm1$, so that $\tanh\left(  \beta Js_{l}\sum_{j}%
A_{lj}s_{j}\right)  =s_{l}\tanh\left(  \beta J\sum_{j}A_{lj}s_{j}\right)  $ we
finally obtain%
\begin{equation}
\frac{w\left(  \mathbf{s}_{l}\rightarrow\mathbf{s}_{l}^{\prime}\right)
}{w\left(  \mathbf{s}_{l}^{\prime}\rightarrow\mathbf{s}_{l}\right)  }%
=\frac{1-s_{l}\tanh\left(  \beta J\sum_{j}A_{lj}s_{j}\right)  }{1+s_{l}%
\tanh\left(  \beta J\sum_{j}A_{lj}s_{j}\right)  } \label{secondo}%
\end{equation}
Therefore, a possible choice for the transition rates for the Ising-Glauber
dynamics on general graph is%

\begin{equation}
w\left(  \mathbf{s}_{l}\rightarrow\mathbf{s}_{l}^{\prime}\right)
=\alpha\left[  1-s_{l}\tanh\left(  \beta J\sum_{j}A_{lj}s_{j}\right)  \right]
\label{terzo}%
\end{equation}
where $\alpha$ is an arbitrary constant. We have chosen $\alpha=1/2$.

\section{Detailed balance condition\label{balance cond}}

Here we show that the transition rate given in Eq. (\ref{transition rates.})
fulfils the detailed balance condition of Markov chains, which requires the
existence of a stationary probability distribution $P_{0}\left(
\mathbf{s}_{l}\right)  $ such that
\begin{equation}
\frac{P_{0}\left(  \mathbf{s}_{l}^{\prime}\right)  }{P_{0}\left(
\mathbf{s}_{l}\right)  }=\frac{w\left(  \mathbf{s}_{l}\rightarrow
\mathbf{s}_{l}^{\prime}\right)  }{w\left(  \mathbf{s}_{l}^{\prime}%
\rightarrow\mathbf{s}_{l}\right)  } \label{enforcign detailed balance}%
\end{equation}
Using Eq. (\ref{transition rates.}) the above condition Eq.
(\ref{enforcign detailed balance}) writes%
\begin{align}
\frac{P_{0}\left(  \mathbf{s}_{l}^{\prime}\right)  }{P_{0}\left(
\mathbf{s}_{l}\right)  }  &  =\frac{1-s_{l}\tanh\left(  \beta J\sum_{h}%
A_{lh}s_{h}\right)  }{1+s_{l}\tanh\left(  \beta J\sum_{h}A_{lh}s_{h}\right)
}\label{quattro}\\
&  \times\frac{\exp\left\{  \beta^{\prime}\left[  \bar{V}\left(
\mathbf{s}_{l}^{\prime}\right)  -\bar{V}\left(  \mathbf{s}_{l}\right)
\right]  \right\}  }{\exp\left\{  \beta^{\prime}\left[  \bar{V}\left(
\mathbf{s}_{l}\right)  -\bar{V}\left(  \mathbf{s}_{l}^{\prime}\right)
\right]  \right\}  }\nonumber
\end{align}
and recalling Eqs. (\ref{detailed balance gluaber}, \ref{secondo}) yields
\begin{equation}
\frac{P_{0}\left(  \mathbf{s}_{l}^{\prime}\right)  }{P_{0}\left(
\mathbf{s}_{l}\right)  }=\frac{\exp\left[  -\beta E\left(  \mathbf{s}%
_{l}^{\prime}\right)  +2\beta^{\prime}\bar{V}\left(  \mathbf{s}_{l}^{\prime
}\right)  \right]  }{\exp\left[  -\beta E\left(  \mathbf{s}_{l}\right)
+2\beta^{\prime}\bar{V}\left(  \mathbf{s}_{l}\right)  \right]  }
\label{cinque}%
\end{equation}
this allows to define the stationary probability distribution%
\begin{equation}
P_{0}\left(  \mathbf{s}_{l}\right)  =\frac{\exp\left[  -\beta E\left(
\mathbf{s}_{l}\right)  +2\beta^{\prime}\bar{V}\left(  \mathbf{s}_{l}\right)
\right]  }{\sum_{k}\exp\left[  -\beta E\left(  \mathbf{s}_{k}\right)
+2\beta^{\prime}\bar{V}\left(  \mathbf{s}_{k}\right)  \right]  } \label{sei}%
\end{equation}
which satisfies the detailed balance condition Eq.
(\ref{enforcign detailed balance}).

\section{The stochastic simulation algorithm\label{Gillespie_alg}}

The stochastic simulation algorithm we use to solve the Markov process
(\ref{Markov chain}) is derived from the one proposed by Gillespie
\cite{Gillespie1, Gillespie2}. We just summarize the main steps of the algorithm:

\begin{enumerate}
\item Choose a random initial state $\mathbf{s}$ of the system

\item Calculate the transition rates $w\left(  \mathbf{s}_{l}\rightarrow
\mathbf{s}_{l}^{\prime}\right)  $

\item Calculate the total rate $w_{T}=\sum_{l}w\left(  \mathbf{s}%
_{l}\rightarrow\mathbf{s}_{l}^{\prime}\right)  $

\item Normalize the transition rates as $\nu_{l}=w\left(  \mathbf{s}%
_{l}\rightarrow\mathbf{s}_{l}^{\prime}\right)  /w_{T}$

\item Construct the cumulative distribution $F\left(  \nu_{l}\right)  $ from
the probability mass function $\nu_{l}$

\item Calculate the time $\Delta t$ to the next opinion flip drawing from an
exponential distribution with mean $1/w_{T}$, i.e. choose a real random number
$0\leq r<1$ from a uniform distribution and set $\Delta t=-w_{T}^{-1}%
\log\left(  r\right)  $.

\item Identify the $k$-th opinion $s_{k}$ which flips from $s_{k}$ to $-s_{k}$
by drawing from a discrete distribution with probability mass function
$\nu_{l}$, i.e. draw a real random number $0\leq s<1$ from a uniform
distribution and choose $k$ so that $F\left(  \nu_{k-1}\right)  \leq
s<F\left(  \nu_{k}\right)  $.

\item Update the state vector and return to step 2 or quit.
\end{enumerate}

\section{The mean-field calculations of the Ising model on a complete
graph.\label{mean-field}}

On a complete graph the total energy of a system of $M$ spins is%
\begin{equation}
E=-\sum_{k<h}J\sigma_{k}\sigma_{h} \label{energy on complete graph}%
\end{equation}
and the average magnetization is $\left\langle \sigma\right\rangle =M^{-1}%
\sum_{k}\left\langle \sigma_{k}\right\rangle $. Using Eq. (\ref{terzo}) the
Ising-Glauber rate becomes%

\begin{equation}
w_{k}=w\left(  \sigma_{k}\rightarrow-\sigma_{k}\right)  =\frac{1}{2}\left[
1-\sigma_{k}\tanh\left(  \beta J\sum_{j\neq k}\sigma_{j}\right)  \right]
\label{rate complete graph}%
\end{equation}
Using Eq. (\ref{Markov chain}) one can easily derive the following equation of
motion for the average magnetization $\left\langle \sigma_{k}\right\rangle $
of the $k$-th site%
\begin{equation}
\frac{d\left\langle \sigma_{k}\right\rangle }{dt}=-2\left\langle w_{k}%
\sigma_{k}\right\rangle =-\left\langle \sigma_{k}\right\rangle +\left\langle
\tanh\left(  \beta J\sum_{j}\sigma_{j}\right)  \right\rangle
\label{equation motion average mag}%
\end{equation}
Assuming that $M$ is large, using $\left\langle \sigma\right\rangle
=M^{-1}\sum_{k}\left\langle \sigma_{k}\right\rangle $, and exploiting the mean
field approach we write $\left\langle \tanh\left[  \beta J\sum_{j}\sigma
_{j}\right]  \right\rangle =\tanh\left[  \beta J\sum_{j}\left\langle
\sigma_{j}\right\rangle \right]  =\tanh\left[  \left\langle \sigma
\right\rangle \beta JM\right]  $, and%
\begin{equation}
\frac{d\left\langle \sigma\right\rangle }{dt}=-\left\langle \sigma
\right\rangle +\tanh\left[  \left\langle \sigma\right\rangle \beta JM\right]
\label{average magnetization equation}%
\end{equation}
The average magnetization $\left\langle \sigma\right\rangle $ at the fixed
point of Eq. (\ref{average magnetization equation}) satisfies the relation%
\begin{equation}
\left\langle \sigma\right\rangle =\tanh\left[  \left\langle \sigma
\right\rangle M\beta J\right]  \label{fixed points}%
\end{equation}
For $M\beta J\leq1$ only the trivial solution $\left\langle \sigma
\right\rangle _{1}=0$ can be found. However, for $M\beta J>1$ other two
solutions $\left\langle \sigma\right\rangle _{2}=-\left\langle \sigma
\right\rangle _{3}>0$ appear which depends on the specific value of $M\beta
J$. In this case $\left\langle \sigma\right\rangle _{1}=0$ becomes unstable.
Thus, the critical point for the phase transition is
\begin{equation}
\left(  \beta J\right)  _{c}=\frac{1}{M} \label{critical point}%
\end{equation}
The above equation is in perfect agreement with the results shown in Fig.
\ref{Figure-criticality}, thus confirming that the decision making process of
the group becomes optimal just when the system changes qualitatively its dynamics.

\end{document}